# The eclipsing binary HS0705+6700 and the search for circumbinary objects

Pulley D*., Faillace G., Smith D., Owen C.

(This paper is to be published in the BAA Journal. Tables and figures are included at the end of this paper)

## Abstract

*HS0705+6700 (also identified as V470 Cam) is a short period (2.3 h) post common envelope detached eclipsing sdB binary system which exhibits transit time variations (TTVs) of a cyclical nature. We report a further 25 timings of light minima and show that our new TTVs support and extend this cyclical pattern to 1.6 periods. We examine possible causes of the observed TTVs and confirm that the presence of a third, and possibly a fourth, body could provide an elegant explanation of these cyclical variations. However other non-circumbinary mechanisms, e.g. Applegate magnetic dynamo effects, will remain possible contenders until sufficient data has been accumulated to demonstrate that the periodicity of the TTVs is time independent.*

## Introduction

HS0705+6700 (hereafter HS0705) is a detached eclipsing binary system that was first investigated by Drechsel et al. in 2001.[1] HS0705 is a member of the HW Vir family of short period binary systems that consist of a very hot subdwarf B type (sdB) star and cool, low mass, main sequence star or brown dwarf. Their compact structure and the large temperature difference between the two components give rise to short and well defined primary eclipses allowing times of minima to be determined with high precision. The sdB component of these systems have masses of ~$0.5M_o$ and consist of a helium burning core with a thin hydrogen envelope and are located at the left hand extremity of the horizontal branch in the H-R diagram.

Various evolutionary scenarios have been proposed for these stars, but a definitive mechanism remains to be established, in particular whether or not binary evolution, as outlined by Paczynski, Webbink and Zorotovic, is a requirement.[2][3][4] These models suggest mass transfer from the primary to secondary component occurs at a rate that cannot be accommodated by the secondary component. This results in material filling first the Roche lobe of the secondary component and then the lobe of the primary resulting in a common envelope enshrouding the binary system. Angular momentum is transferred from the binary system to the common envelope bringing the binary pair closer together and resulting in a short binary period of typically between 2 and 3 hours. Eventually the common envelope has sufficient energy to be violently ejected from the system.

Zorotovic et al provided an overview of thirteen of these systems in 2013.[5] Interestingly in five of the thirteen systems eclipse TTVs have been interpreted as showing the presence of low mass circumbinary objects e.g. brown dwarfs, massive planets etc. While some of the cases remain unclear, the evidence provided for a third body, based upon the cyclical eclipse TTVs, increases. If such bodies do exist then they must have either survived the energetic common envelope ejection process or have formed during the short period since the common envelope was ejected.

In this paper we review the transit time variations exhibited by one such system, HS0705, over the last thirteen years and explore whether there is conclusive evidence that these TTVs can predict the presence of a third body. We first look at the documented historical measurements made on HS0705 and their subsequent analyses since 2001. We then discuss our 25 new measurements taken between late 2013 and early 2014 before completing a period and transit time variation analysis, using both unweighted and weighted data, on all known times of minimum. Finally we discuss the possible causes of the observed TTVs and the possibility of the presence of a third body, and potentially a fourth body, before presenting our conclusions.

* Contact david@davidpulley.co.uk

## Historical Measurements and Analysis

HS0705 was identified as an sdB star from the Hamburg Schmidt Quasar Survey and confirmed by Drechsel et al. as a short period eclipsing binary system after a follow up monitoring programme with the 2.5m Nordic Optical Telescope.[1] They produced the first linear ephemeris in 2001 and they were also the first to derive a radial velocity curve from the optical spectra and first to obtain the system parameters using a modified Wilson Devinney code. Drechsel et al. analyses determined that HS0705 was a detached eclipsing binary system with an orbital period of approximately 2.3 hours, a mass ratio of q=0.28 and an inclination of 84.4 degrees. The effective temperature of the primary sdB component was calculated to be 28,800 ±900K. They concluded that the M dwarf companion did not contribute to the optical light of the system except for the light reflected from the hemisphere facing the sdB star.

Niarchos et al. added further data in 2003 as did Qian et al. in 2009[6][7]. Qian et al. also noted that the time of minima showed a cyclical change suggesting this could be attributed to the presence of a third body in a circular orbit around the binary system with a period of 7.15 years. In 2010 Qian et al. extended his argument for the presence of a third body with additional data points and derived a quadratic ephemeris with a superimposed sinusoidal variation with period 15.7 years.[8] We find this period is incompatible with his Eq. 2 and believe the period should be corrected to 7.15 years. Camurdan et al. added further data in 2012 but disagreed with Qian et al.'s prediction of a quadratic ephemeris and reverted to a linear ephemeris with a superimposed cyclical variation corresponding to an elliptical orbit using the methodology of Irwin. [9][10] Niarchos et al. and Camurdan et al. both performed light curve analysis and computed similar system parameters to Drechsel et al.[1][6][9]

Beuermann et al. provided additional data in 2012 which included an early mean eclipse timing from archival data of the Northern Sky Variability Survey [11][12] An overview of this survey is provided by Wozniak et al. [13]. Beuermann et al. derived an elliptical orbit with an underlying linear ephemeris. Qian et al. added a further 78 data sets in 2013 and compared a potential third body elliptical orbit superimposed on (i) a linear ephemeris and (ii) a quadratic ephemeris[14]. Favouring the quadratic fit for its lower residuals, Qian et al. suggested that the apparent quadratic term may result from a long period fourth body orbiting the system. Two further data sets are recorded in IBVS 5599 and 5796 [15][16]

## Observations and photometric reduction

We report 25 new observations of HS0705 made between September 2013 and May 2014 using Sierra Stars and iTelescope robotic telescopes and South Stoke Observatory. To minimise the effects of differing atmospheric extinction all observations were made at altitudes of greater than $40^0$. All images were calibrated using dark, flat and bias frames and then analysed with Maxim DL software employing aperture photometry. The target was compared with four reference stars and the average magnitude from the four measurements was derived. The reference stars were taken from the American Association of Variable Stars Observer (AAVSO) Photometric All sky Survey (APASS catalogue). Details of the telescopes, target star and reference stars used can be found in Tables 1 and 2.

## Period Analysis and Timing Variations

*Period analysis*
Table 3 lists the data we collected between September 2013 and May 2014. Timings were captured in JD and were converted to HJD and finally to BJD. [17][18]  We analysed our data and determined the minima using the Kwee and van Woerden procedure implemented through two software packages, Peranso ver. 2.51 and Minima25c ver. 2.3. [19][20][21]   We took the arithmetic mean of these two calculated minima which showed a 1-sigma spread of 2.1s. Our data combined with all known historic measurements comprise 214 data sets which can be found on the BAA Variable Star Section website (menu option Articles/Observing).[22]  All historic times of minima have been converted to BJD.

*Linear ephemeris:*  Drechsel et al. found HS0705 to be a short period (~ 2.3$^h$) detached sdB binary system.[1]   Unlike cataclysmic variables, sdB binary systems show no indication of a period gap between 2$^h$ and 3$^h$ e.g. Morales-Rueda et al.[23]  We examined the data assuming a linear ephemeris as might be expected from a detached system. We removed data point E = 30149 from this analysis since its computed (O - C) value deviated by more than 5 standard deviations from expected values.   This was also removed by Qian et al. in their analysis. [7]  Our computed linear ephemeris is:

$$BJD = 2451822.76006(15) + 0.095646671(4)$$

The standard error of the residuals is 7.77 x 10$^{-4}$ and the residuals are shown in Fig. 1.  The distribution of residuals indicates a possible quasi periodic variation superimposed on the linear ephemeris which has been interpreted by many investigators as indicative of a light travel time effect of a third body orbiting the binary system.[7][8][9][11][14]

*Periodic ephemeris:*  Qian et al. were the first to suggest that the apparent cyclical nature of the residuals was the effect of a third body orbiting the binary system and could be accounted for by adding a simple sine term to the linear ephemeris.[7]   Camurdan et al. generalised this approach introducing an elliptical orbit based upon the equations of Irwin and writing the ephemeris in the form: [9][10]

$$BJD = T_0 + P \times E + \tau_3$$

where $T_0$ is the epoch, P is the binary period, E is the binary period number and $\tau_3$ is the light travel time effect given by:

$$\tau_3 = \frac{a_{12} \sin i}{c}\left[(1-e^2)\frac{\sin(v+\omega)}{1+e\cos v} + \sin\omega\right]$$

where $a_{12}\sin i$ is the projected binary semi major axis, c is the speed of light, e the orbital eccentricity, $\omega$ the longitude of periastron passage and *v* the true anomaly.  This can be rewritten in terms of the mean anomaly and eccentric anomaly e.g. Qian et al.[14]  Using this formulation we found the revised linear ephemeris for unweighted data as:

$$BJD = 2451822.76158(2) + 0.095646632(1) + \tau_3$$

We disregarded three data sets with residuals larger than 5 standard deviations from expected values (E = 1409, 30149, 30150) as did Drechsel et al. and Qian et al.[1][7]   The revised ephemeris had a reduced standard error of residuals of $21.53 \times 10^{-5}$. We calculated the orbital period of the potential third body as 8.44 years with an eccentricity of 0.22. The (O - C) residuals are shown in Fig. 2a and the calculated orbital parameters are listed in Table 4.

***Quadratic ephemeris and a potential fourth body:***  Long term period changes of HS0705 have been investigated by Qian et al. by introducing a quadratic term into the ephemeris of this binary system.[8][14]   We investigated this and determined a new ephemeris for unweighted data as:

$$BJD = 2451822.76155(5) + 0.0956466090(42) \times E + 5.49(86) \times 10^{-13} \times E^2 + \tau_3 \quad (1)$$

where $\tau_3$ is the cyclical light travel time effect whose parameters are listed in Table 4 and the (O - C) residuals are shown in Fig. 2b.  This solution reduced the residuals to $20.60 \times 10^{-5}$ providing a marginally better fit to the data.  Interpreting the light travel time effect as caused by a third body we find a slightly increased period of 8.55 years but a low orbital eccentricity of 0.03. The positive coefficient of the quadratic term suggests a long term period increase and in agreement with Qian et al.[14]

***Binning and weighting of data:***  Improvement in the apparent precision of measurement can often be achieved by binning or weighting data.  Binning groups of data, usually on a timeline basis, replaces the binned data points by a single "average" data point. Binning will always result in a loss of information but, if the errors are randomly distributed, what remains is potentially more precise.  Weighting of data is a process that gives greater emphasis to some data sets over other data sets. This emphasis should be based upon measurement accuracy i.e. more accurate measurements are given greater emphasis over less accurate measurements. Effective application of weighting requires a knowledge of the true measurement uncertainty, which can be problematic when assessing historical data. Both binning and weighting can mask fine structure contained within the data.

Various methodologies of binning and weighting have been reported.  Beuermann et al. binned HS0705 data into weekly or monthly buckets dependent upon the (O - C) scatter whereas  Qian et al. weighted his data using *"the reciprocal of the square of the error"*.[11][14]   Beuermann et al., in analysing minima of NN Serpentis,[24] *"weights them by their statistical errors"*.  Neither Beuermann et al. nor Qian et al. clearly state which errors they use, however the inference is that the statistical errors are taken from the minimum calculation of Kwee and van Woerden.[19]

We explored weighting using several weighting methodologies and found that the derived periodic parameters depended significantly upon which weighting methodology used.  Using a weighting methodology similar to Qian et al. our results yield, for a linear ephemeris, a period and eccentricity of 9.31 years and 0.17, respectively, whereas Qian et al. reports 9.53 years and 0.22 and Beuermann et al., using binning, reports 8.41 years and 0.38 eccentricity.[11][14]  We found that weighting the quadratic ephemeris reduced the period to 8.73 years and increased the eccentricity to 0.22 which compares with Qian et al. of 8.87 years and 0.19.[14]   Our calculated residuals were reduced from $8.71 \times 10^{-5}$ to  $7.86 \times 10^{-5}$.  The orbital parameters we determined using weighting are given in Table 4 and the (O - C) residuals shown in Fig. 2c and 2d.  Fig. 3 shows the cyclical variation for a quadratic ephemeris for both unweighted and weighted residuals.

## Discussion

In the preceding analysis the (O - C) residuals of HS0705 suggest that the observed transit time variations (TTVs) comprise of a complex combination of sinusoidal variations superimposed upon a potential longer term monotonic change in binary period.  Possible causes of these two  effects are discussed in this section. First we consider ***periodic variations*** in the (O - C) curves. These variations may be composed of constant and/or quasi periodic sinusoids.

***Magnetic coupling:*** Applegate and others have postulated that binary systems containing a star with a convective envelope can exhibit quasi sinusoidal binary period modulation.[25] A magnetic dynamo effect is thought to occur in the convective envelope or at the convective envelope/radiative core interface. Applegate showed that sub-surface magnetic fields of a few kilogauss can produce a torque that slowly changes the star's angular momentum whilst maintaining hydrostatic equilibrium. This in turn affects the star's oblateness and is transmitted to the binary's orbital parameters by its gravitational quadrupole moment. The model proposed by Applegate considers a thin shell of less than 10% of the convective star's mass being spun up to transfer the angular momentum. This mechanism requires an energy budget that has to be supplied during the modulation period of the system by a variation of the convective star luminosity of <10% and a differentially rotating shell to core angular velocity at the 0.01 level.

The Applegate mechanism assumes that the energy required to modulate the binary period comes from a variation in luminosity of the convective secondary star. Observation of this luminosity variation of HS0705 has not been recorded. This is possible if the thermal time constant of the secondary envelope is several orders of magnitude greater that the modulation period of the binary system, see for example Watson et al.[26]

Earlier publications have suggested that the Applegate mechanism is unlikely to be sufficiently strong to produce the cyclical period variations observed in HS0705.[7][11] Qian et al. has analysed HW Vir [27], an sdB binary system, and shown that the energy budget could not be met; other authors have called upon the analysis of NN Ser, a hydrogen rich white dwarf and an M4 dwarf star binary system, (see for example Brinkworth et al.,[28]) which again shows an insufficient energy budget to explain the quasi periodic (O - C) variations. Here we consider the Applegate mechanism specifically applied to HS0705. Constraints on the shell mass and shell angular momentum are given by the corrected Applegate Eq. 29 and shown in Eq. 2 below.[29][30]...

$$\frac{M_s}{M}\frac{\Delta\Omega}{\Omega} = \frac{GM}{2R^3}\left(\frac{a}{R}\right)^2\left(\frac{P}{2\pi}\right)^2\frac{\Delta P}{P} \qquad (2)$$

where $M_s$, $M$ and $R$ are the convective star's shell mass, total mass and radius. Drechsel et al. determined the convective stars mass and radius with respect to solar parameters as 0.134 and 0.186.[1] For HS0705 the ratio of the binary separation to the convective star radius, a/R, is 4.03 and a binary period, P, equal to 8264s. The (O - C) semi amplitude of the quasi sinusoid is taken as 95s (see Table 4) and the orbital period modulation, $\Delta P/P$, computed from Applegate Eq. 38, is $2.16 \times 10^{-6}$. From Eq. 2 we get the differential angular velocity, $\Delta\Omega/\Omega$, of $2.49 \times 10^{-3}$. Application of this value to Applegate Eq.28 provides an estimate of the energy required to transfer the angular momentum from the core to the outer shell of the convective star as $2.13 \times 10^{40}$ ergs. The energy budget provided by the convective star, assuming a maximum luminosity variation of 0.1, can be found from Applegate Eq. 30 as $7.4 \times 10^{37}$ ergs and a subsurface magnetic field strength (Applegate Eq.33) required of 24.7kG. This field is an order of magnitude greater than is typically expected from the Applegate mechanism. The fusion energy generated within the convective star is also two orders of magnitude too small to drive the levels of quasi sinusoidal variation seen in this binary system.

For the Applegate process to provide a viable mechanism to produce the observed quasi sinusoidal variation, other energy sources need to be identified. A more efficient mechanism to transfer angular momentum has been proposed by Lanza et al. (1998) modifying the Applegate mechanism whereby magnetic energy interchanges with rotational kinetic energy.[31] It is difficult to assess the impact of Lanza's approach on the magnetic dynamo effect except that it should reduce the energy required to effect the period modulation. Whether this is sufficient to reduce the energy requirement by two orders of magnitude remains an unanswered question.

Although our analysis indicates that the intrinsic luminosity of HS0705's fully convective secondary is insufficient to power an Applegate type mechanism, the secondary star is close (~0.8$R_0$) to its very hot sdB primary. The secondary star will be irradiated by the primary receiving approximately 300 times more energy than it can generate from its own internal nuclear processes. Over the modulation period the intercepted energy is of the order 6 x $10^{41}$ergs thus exceeding by more than an order of magnitude the energy requirement needed to drive the Applegate mechanism. Brinkworth et al. suggests that most of the incident energy will be reflected from the surface layers of the fully convective star and little will penetrate the high opacity layers to a depth required to generate an Applegate type mechanism.[28] However the energy interaction between the two stars is governed by complex physical processes that are not well understood. Computer modelling by Claret and others suggests that the bolometric albedo of the fully convective secondary is much less than the unity value that was assumed by Brinkworth et al.[32] Furthermore the incident energy can alter the structure of the secondary star and some will perturb the deeper layers within this body as described by Ruciński (1969) and Brett et al. (1993).[33] [34] Whether this is sufficient to drive an Applegate type mechanism remains an open question.

***Periastron precession:*** The precession of the periastron of the binary orbit can give rise to TTVs originating from both classical Newtonian mechanics and General Relativity (GR). Such variation would be expected to show a constant period from cycle to cycle. The effects manifesting from Newtonian mechanics are related to gravitational interactions induced by other massive bodies orbiting the binary system and the oblateness of the stars within the system. These effects are usually measured on a secular timescale.

The precession period, $P_p$, due to GR can be calculated from...

$$P_p = \frac{c^2(1-e^2)}{3(2\pi GM)^{2/3}} P^{5/3} \qquad (3)$$

where e is the orbital eccentricity, P is the binary period, M is the total mass of the system and c and G are the speed of light and Universal Gravitational constant. Eq. 3 can be derived by combining the GR expression for perihelion advance with Kepler's Third Law, see for example Pal et al.[35] A short precession period will occur for close in, short period, binary systems but with these systems orbits will be expected to be of low eccentricity. For HS0705 the precession period calculated from Eq. 3 is approximately 50 years and significantly greater than the 8 to 9 year period observed for the quasi periodic residuals calculated from its quadratic ephemeris and shown in Fig. 3.

***Light travel time effects:*** A third body orbiting a binary system will produce transit time variations that exhibit a periodic pattern similar to those shown in Fig. 3. The orbiting third body causes the barycentre of the binary pair to move closer to or further away from an Earth bound observer and hence decreases or increases the respective light travel time from the system. As the barycentre of the binary pair move closer to the Earth, primary eclipses will occur earlier than expected. Conversely as the barycentre of the binary pair move away from the Earth the eclipses will occur later than expected. This will be repeated for each orbit of the third body.

We can discriminate between cyclical TTVs produced by a third body and those produced by a magnetic dynamo effect (e.g. Applegate) by observing the constancy of the observed cyclical period of the TTVs. A third orbiting body will generate a cyclic TTV of almost constant period whereas the cyclical TTVs of magnetic dynamo will be quasi periodic. The period of the magnetic dynamo will vary continuously in a similar fashion to the variable 11 year sunspot cycle of our Sun. Thus long term observations of the stability of cyclical TTVs provides a useful methodology for discriminating between the presence of a third body and other non-circumbinary effects e.g. magnetic dynamo mechanisms.

If the cyclical TTVs observed in HS0705 are produced by a third body we are able to put constraints upon the mass of the third body and its orbital parameters. The mass function of the third body is given by...

$$f(M_3) = \frac{(M_3 sini)^3}{(M_1+M_2+M_3)^2} = \frac{4\pi^2 a_{12}^3 sin^3 i}{G P_3^2}$$

where $M_1$, $M_2$ are the binary star masses and $M_3$ of the mass of the third body; $i$ is the orbital inclination relative to the plane of the sky; $P_3$ is the orbital period of the third body and $a_{12}sini$, the projected semi-major axis of the binary calculated from the light travel time effect. Table 5 summarises the third body parameters, calculated from the light travel time effect, for both unweighted and weighted linear and quadratic ephemeris. These results are compared with Qian et al. (2013) for their weighted linear and quadratic ephemeris.

Whilst there are some similarities between the results of Qian et al. (2013) and the results from this paper, Table 5 does indicate that there is a wide spectrum of results which are dependent upon the choice of methodology (unweighted vs weighted and linear vs quadratic) used to determine the ephemeris. For example this analysis (Table 5) shows the orbital period and mass of the third body, in terms of Jupiter's mass, are nominally 9.0 years and 35.2$M_J$ respectively but extend over +/-0.5 years and +/-5.5$M_J$.

The mass of the third body is inversely proportional to the sine of the orbital inclination, relative to the plane of the sky, giving a minimum mass, for an orbital inclination of $90^0$, of between 29.6$M_J$ and 40.7$M_J$. The mass of the third body will increase as the orbital inclination decreases indicating this object to be a potential brown dwarf or possible stellar object. Assuming a transition mass from brown dwarf to stellar object of 75$M_J$ and a transition mass from brown dwarf to planet of 13$M_J$ our analysis indicates that the third body will be a brown dwarf if the orbital inclination is greater than ~ $33^0$. The orbital radius of the third body will be greater than ~ 3.6AU.

Longer term *monotonic changes* in the (O - C) curves are now considered. By including the quadratic term in the ephemeris of Eq. 1 we are able to minimise the (O - C) residuals and provide the best fit to the data. Qian et al. (2010) was first to introduce the quadratic term with a negative coefficient suggesting there was a period decrease.[8] Subsequently Qian et al. (2013) revised their analysis with a positive coefficient quadratic indicating period growth.[14] Our analysis confirms a positive coefficient indicating a long term growth in the binary period. Drechsel et al. showed that HS0705 is a detached binary system thus binary period changes by mass transfer through the Roche lobe are not possible .[1] Some causes of monotonic period changes are reviewed below.

***Magnetic stellar wind breaking and gravitational waves:*** Our revised ephemeris, Eq. 1, assigns a positive coefficient to the quadratic term indicating a period increase with time. Angular momentum loss through either magnetic stellar wind breaking or gravitational wave radiation will reduce the binary period and we conclude that neither of these mechanisms can explain the overall underlying observed period increase. These effects may be present, having a role to play over the very long term evolution of HS0705, but masked by other period increase mechanisms.

*Proper motion:* Reorientation of the binary system with respect to the Earth due to its proper motion will cause the period to appear to change due to its changing distance and consequential change in light travel time (see for example Rafikov).[36] The magnitude of this change will depend on the relative velocity and distance of the object from the observer. These parameters remain unknown for HS0705 but if we assume a nominal value of 30kms$^{-1}$ for the transverse velocity and a nominal distance of 200pc (these values are not known but what may be thought as typical for this exercise) this approximates to a change of ~1.5 x 10$^{-12}$dyr$^{-1}$. This is three orders of magnitude less than predicted by the quadratic ephemeris given here and also by Qian et al.[14] Although unlikely the proper motion effects may come into contention if higher velocities and smaller distances are appropriate for HS0705.

*Fourth body:* Qian et al. (2010 and 2013) has suggested that the quadratic term in their ephemeris for HS0705 might be caused by a fourth body orbiting the binary pair.[8][14] The positive coefficient of the quadratic term in Eq. 1 and the detached nature of the binary system eliminates some TTV mechanisms, e.g. mass transfer, gravitational waves, magnetic stellar wind breaking, adding strength to the fourth body concept. The orbital parameters of a fourth body can be deduced by including a second light travel time effect, $\tau_4$, in Eq. 1. Beuermann et al. used this approach to infer two planets orbiting NN Ser.[24] Interestingly Qian et al. (2010) derived a negative coefficient for the quadratic term. Their ephemeris did assume a circular orbit but is in marked contrast with Qian et al. (2013) and our work that derives a similar positive quadratic term. This is in need of further investigation but may support the notion of the fourth body.[8][14]

Justifying the presence of a fourth body will require precise and accurate observations to be taken over a sufficiently long time period to eliminate other causes of TTVs. The presence of a fourth body can be tested by performing orbital stability analysis and assessing the longevity of the calculated orbits using computer modelling techniques suggested by Horner et al.[37] As yet there is no firm evidence for a fourth body orbiting HS0705.

## Conclusions and further work

In the previous section we discussed some of the ways transit time variations may arise in binary systems and noted that in HS0705 these variations appeared to be of a periodic nature with an underlying monotonic growth. Interpreting these changes can be challenging. The apparent periodic element of the TTVs can be explained by the presence of a third body, possibly a brown dwarf or stellar object, with orbital period of approximately 9 years and with a mass greater than 29$M_J$. It was also noted that the derived parameters were dependent upon the weighting methodology employed which may be a reflection of the accuracy of the recorded times of minima.

With the growing preponderance of reported exo-planets and the elegance of a third body solution, it is easy to dismiss other possible explanations such as magnetic dynamo effects within the fully convective secondary star. Although Applegate's proposed nuclear energy source within the secondary may be inadequate to drive such changes, other more efficient angular momentum change mechanisms have been suggested and the presence of the very hot primary sdB star irradiating the nearby secondary could potentially supply sufficient energy to drive this mechanism.

The quadratic term in the ephemeris minimises the (O - C) residuals and drives a monotonic increase in binary period. Many mechanisms driving such change either predict a period decrease (e.g. magnetic stellar wind breaking or gravitational waves) or are inappropriate (e.g. mass transfer) or too small to have an impact (e.g. proper motion). Interestingly Qian et al. suggested that the apparent quadratic term may be part of a much longer period change introduced by a fourth body. There is as yet no firm evidence for this and, if this observed change is part of a longer periodic cycle, then a magnetic dynamo mechanism may be responsible as suggested by Applegate in the case of Algol. Orbital stability analysis may assist in constraining parameters of such multi-bodied systems.

Whilst the presence of a third body is the strongest contender for explaining the observed TTVs, other non-circumbinary mechanisms still provide viable options. Key to resolving these issues will be to determine whether the observed cyclical TTVs have a constant period, as expected from a third body, or a variable period typical of an Applegate type mechanism. Further evidence for a third body may be gained from detection of the reflex radial velocity of the sdB star due to the third body. It is estimated that this will be of the order of 0.5kms$^{-1}$ and superimposed upon the binary reflex velocity measured by Drechsel et al. of ~ 85.8kms$^{-1}$. However this may well fall within measurement uncertainty.[1] Detection of system brightening at superior conjunction, caused by the reflected light off the third body is conceivable; Camurdan et al. assumed a 4% brightening at 0.25 phase in their light curve analysis.[9] The detection of an eclipse of the binary pair by the third body is also a remote possibility.

## Acknowledgements

The authors would like to thank Dr. Iain Chapman, University of Durham, for his suggestions on regression analysis of circular orbits and Dr. David Boyd, BAA, and Dr. Robert Smith, University of Sussex, for their thoughts on weighting of data. We are also grateful to the referees who provided constructive criticism of this document. This work was supported by the BAA who funded many of our observations taken in 2013/14 and made use of the APASS database maintained on the AAVSO website.

| Observatory | Telescope | Instrumentation |
|---|---|---|
| Sierra Stars Observatory Markleeville, Ca, USA http://sierrastars.com/gp/SSO/SSO-CA.aspx | 0.61m $f_L$ 6100mm | Finger Lakes Inst. ProLine camera 3056 x 3056 pixels FOV 21 x 21 arcmin |
| Sierra Stars Observatory Mt Lemmon, Az, USA http://sierrastars.com/gp/MLSC32/MLSC32.aspx | 0.81m $f_L$ 5670mm | SBIG STX KAF-16803 camera 4096 x 4096 pixels FOV 22.5 x 22.5 arcmin |
| iTelescope T21 Mayhill, NM, USA http://www.itelescope.net/telescope-t21/ | 0.43m $f_L$ 2920mm | Finger lakes Inst. PL6303E camera 3072 x 2048 pixels FOV 49.2 x 32.8 arcmin |
| South Stoke Observatory Oxfordshire, England | 0.28m $f_L$ 1400mm Celestron | SBIG STL-1100 camera 512 x 512 pixels FOV 25.1 x 25.1 arcmin |

Table 1: Telescopes and instrumentation used for the measurements reported in this paper

| Catalogue No. | Tar/Ref Star No. | RA (J2000) | DEC (J2000) | Johnson Filters V | B | Sloan r' |
|---|---|---|---|---|---|---|
| HS0705+6700 V470 Cam GSC 4123-00265 | Target Star | 07:10:42.42 | +66:55:43.6 | | | |
| GSC 4123:1037 | Ref 1 | 07:11:21.47 | 67:00:57.55 | 13.637 | 14.141 | 13.514 |
| GSC 4123:410 | Ref 2 | 07:09:10.50 | 66:58:51.25 | 13.435 | 14.172 | 13.190 |
| GSC 4123:171 | Ref 3 | 07:09:24.09 | 67:01:45.42 | 13.798 | 14.230 | 13.685 |
| GSC 4123:561 | Ref 4 | 07:10:24.68 | 66:56:18.49 | 13.842 | 14.284 | 13.757 |

Table 2: Coordinates of target and reference stars and and APASS magnitudes of reference stars used for photometry

| JD (+2400000) | HJD (+2400000) | BJD (+2400000) | E | Minima Type | Error (d) | Error (s) | Filter | Telescope |
|---|---|---|---|---|---|---|---|---|
| 56540.916510 | 56540.914645 | 56540.915412 | 49329.0 | I | 0.000071 | 6.09 | V | iTelescope T21 New Mexico |
| 56566.930550 | 56566.930455 | 56566.931222 | 49601.0 | I | 0.000081 | 7.00 | V | iTelescope T21 New Mexico |
| 56574.868787 | 56574.869257 | 56574.870024 | 49684.0 | I | 0.000079 | 6.78 | V | iTelescope T21 New Mexico |
| 56605.951988 | 56605.954512 | 56605.955279 | 50009.0 | I | 0.000024 | 2.07 | G Astrodon | SSON Mt Lemmon |
| 56626.897355 | 56626.900896 | 56626.901664 | 50228.0 | I | 0.000109 | 9.37 | V Johnson | SSON Markleeville |
| 56629.575214 | 56629.578854 | 56629.579622 | 50256.0 | I | 0.000051 | 4.36 | V | iTelescope T7 Nerpio |
| 56631.392758 | 56631.396460 | 56631.397227 | 50275.0 | I | 0.000051 | 4.36 | R | South Stoke |
| 56631.440490 | 56631.444194 | 56631.444961 | 50275.5 | II | 0.000116 | 10.02 | R | South Stoke |
| 56631.488214 | 56631.491919 | 56631.492687 | 50276.0 | I | 0.000097 | 8.34 | R | South Stoke |
| 56655.304247 | 56655.308391 | 56655.309159 | 50525.0 | I | 0.000173 | 14.90 | R | South Stoke |
| 56655.351047 | 56655.355193 | 56655.355960 | 50525.5 | II | 0.000180 | 15.51 | R | South Stoke |
| 56655.399667 | 56655.403812 | 56655.404580 | 50526.0 | I | 0.000066 | 5.66 | R | South Stoke |
| 56655.447632 | 56655.451778 | 56655.452545 | 50526.5 | II | 0.000170 | 14.64 | R | South Stoke |
| 56655.495084 | 56655.499229 | 56655.499996 | 50527.0 | I | 0.000063 | 5.44 | R | South Stoke |
| 56662.812451 | 56662.816583 | 56662.817351 | 50663.5 | II | 0.000143 | 12.31 | B Johnson | SSON Markleeville |
| 56715.371974 | 56715.374145 | 56715.374913 | 51153.0 | I | 0.000060 | 5.18 | Sloan r' | South Stoke |
| 56718.337060 | 56718.339044 | 56718.339812 | 51184.0 | I | 0.000051 | 4.36 | Sloan r' | South Stoke |
| 56723.789332 | 56723.790960 | 56723.791728 | 51241.0 | I | 0.000019 | 1.64 | V | SSON Markleeville |
| 56739.763577 | 56739.764096 | 56739.764864 | 51408.0 | I | 0.000022 | 1.90 | G Astrodon | SSON Mt Lemmon |
| 56740.338146 | 56740.338624 | 56740.339392 | 51414.0 | I | 0.000062 | 5.31 | Sloan r' | South Stoke |
| 56740.433131 | 56740.433602 | 56740.434370 | 51415.0 | I | 0.000105 | 9.03 | Sloan r' | South Stoke |
| 56740.719933 | 56740.720383 | 56740.721152 | 51418.0 | I | 0.000088 | 7.56 | B Johnson | SSON Markleeville |
| 56775.776639 | 56775.774680 | 56775.775448 | 51784.5 | II | 0.000130 | 11.23 | V | SSON Markleeville |
| 56777.737658 | 56777.735578 | 56777.736346 | 51805.0 | I | 0.000036 | 3.07 | B Johnson | SSON Markleeville |
| 56779.746431 | 56779.744230 | 56779.744999 | 51826.0 | I | 0.000039 | 3.33 | V | SSON Markleeville |

Table 3: Primary and secondary eclipse timings taken by the authors between September 2013 and May 2014

| | | Elliptical Linear (Unweighted) | Elliptical Quad (Unweighted) | Elliptical Linear (Weighted) | Elliptical Quad (Weighted) |
|---|---|---|---|---|---|
| Epoch | (d) | 2451822.76158(5) | 2451822.76155(5) | 2451822.76120(2) | 2451822.76148(2) |
| Binary Period | (d) | 0.095646632(1) | 0.0956466090(42) | 0.0956466400(4) | 0.0956466012(16) |
| Rate of Change | (d/yr) | --- | $4.19(66) \times 10^{-9}$ | ---- | $5.82(18) \times 10^{-9}$ |
| Eccentricity | | 0.22(1) | 0.03(1) | 0.17(1) | 0.22(1) |
| Longitude of Periatron | (deg) | 327.70 | 366.84 | 325.03 | 336.09 |
| Periastron Passage | (d) | 2450748(1) | 2449484(1) | 2451990(1) | 2452286(1) |
| Semi-amplitude | (d) | 0.00123 | 0.00102 | 0.00110 | 0.00093 |
| Semi Amplitude | (s) | 106.37 | 88.3 | 95.4 | 80.1 |
| Orbital Period | (yr) | 8.44(1) | 8.55(1) | 9.31(1) | 8.73(1) |
| Projected semi major axis ($a_{12}\sin i$) | (au) | 0.2127 | 0.1767 | 0.1908 | 0.1602 |
| std error of resduals | | 0.0002153 | 0.0002060 | 0.0000871 | 0.0000786 |

Table 4: Orbital parameters of the third body for linear and quadratic ephemeris and unweighted and weighted data

| | | This Paper | | | | Qian et al 2013 | |
|---|---|---|---|---|---|---|---|
| | | Linear UnWeighted | Linear Weighted | Quadratic UnWeighted | Quadratic Weighted | Linear Weighted | Quadratic Weighted |
| Period, $P_3$ | yrs | 8.44 | 9.31 | 8.55 | 8.73 | 9.53 | 8.87 |
| LITE semi amplitude | s | 106.4 | 95.4 | 88.3 | 80.1 | 96.25 | 87.44 |
| Mass Function | $M_0$ | 1.36E-04 | 8.06E-05 | 7.58E-05 | 5.42E-05 | 7.90E-05 | 6.84E-05 |
| $M_3\sin i$ | $M_0$ | 0.0388 | 0.0324 | 0.0317 | 0.0283 | 0.0322 | 0.0306 |
| $M_3\sin i$ | $M_J$ | 40.7 | 33.9 | 33.2 | 29.6 | 33.7 | 32.1 |
| $a_{1,2}\sin i$ | au | 0.213 | 0.191 | 0.176 | 0.160 | 0.192 | 0.175 |
| $a_{12,3}\sin i$ | au | 3.58 | 3.82 | 3.61 | 3.66 | 3.88 | 3.70 |

Table 5: Orbital properties of a potential third body. The results from this analysis are shown under 'This Paper' and compared with Qian et al (2013). Qian reports the orbital distance as the difference between $a_{12,3}\sin i$ and $a_{1,2}\sin i$.

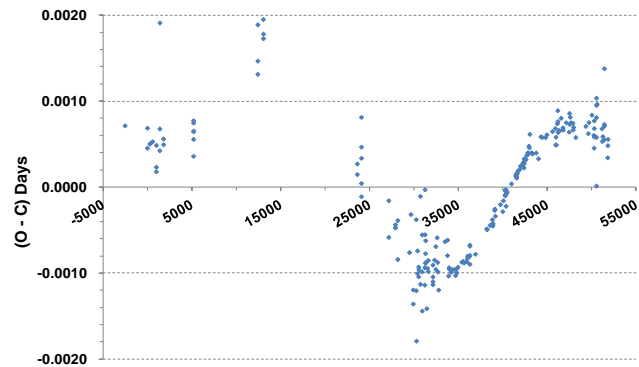

Fig. 1 (O - C) residuals computed from the linear ephemeris. Data point E = 30149 has been removed

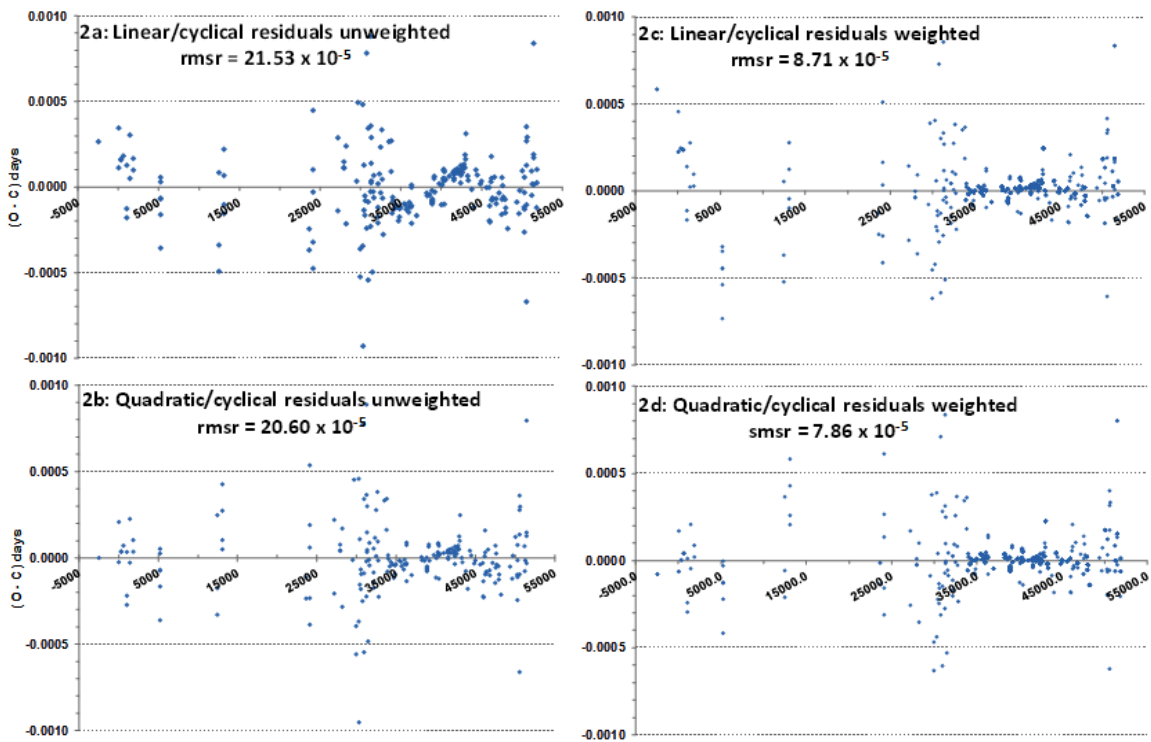

Fig 2. Plot of linear and quadratic residuals with superimposed elliptical orbit light time effects. Figs. 2a and 2b for non weighted data and Figs. 2c and 2d for weighted data

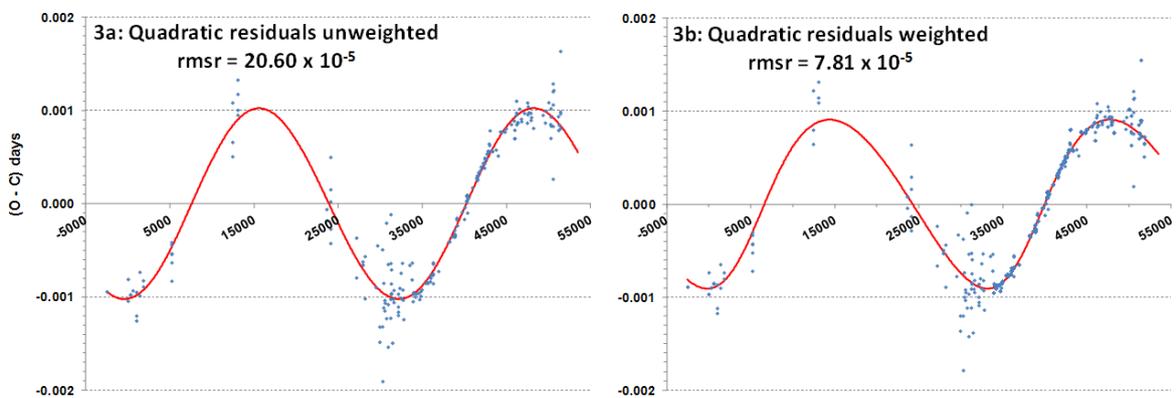

Fig. 3. Ellipsoidal fit to the quadratic ephemeris for weighted and non weighted residuals.